\documentclass[12pt]{article}

\usepackage{amssymb}
\usepackage{amsxtra}
\usepackage{amsmath}
\usepackage{amsfonts}
\usepackage{CJK}
\usepackage[titletoc]{appendix}
\usepackage{graphicx}

\numberwithin{equation}{section}

\newtheorem{thm}{Theorem}[section]

\newtheorem{prop}[thm]{Proposition}

\newtheorem{ass}[thm]{Assumption}

\newtheorem{lem}[thm]{Lemma}

\newtheorem{rem}[thm]{Remark}

\newcommand{\eqa}{\begin{eqnarray}}
\newcommand{\eeqa}{\end{eqnarray}}
\newcommand{\beq}{\begin{equation}}
\newcommand{\eeq}{\end{equation}}
\newcommand{\nn}{\nonumber}

\allowdisplaybreaks

\begin{document}
\title{ { The initial-boundary value problems for the coupled derivative nonlinear Schr\"{o}dinger equations on the half-line }
\thanks{\footnotesize {The work was partially supported by the National Natural Science Foundation of
China under Grant Nos. 11271008, 61072147, 11601055.}}}

\author{{\footnotesize { Bei-bei Hu$^{1,2}$ , Tie-cheng Xia$^{2,}$\thanks{Corresponding author. E-mails: hu\_chzu@163.com(B.-b. Hu), xiatc@shu.edu.cn(T.-c. Xia), zhangningsdust@126.com(N. Zhang)} and Ning Zhang$^{2,3}$ }}\\
{\footnotesize { {\it$^{1}$School of Mathematicas and Finance, Chuzhou University, Anhui, 239000, China}}}\\
{\footnotesize { \it $^{2}$Department of Mathematics, Shanghai University, Shanghai 200444, China}}\\
{\footnotesize { \it $^{3}$Department of Basical Courses, Shandong University of Science and Technology, Taian, 271019, China}}}
\date{\small }\maketitle

\textbf{Abstract}: The unified transform method is used to analyze the
initial-boundary value problem for the coupled derivative nonlinear
Schr\"{o}dinger(CDNLS) equations on the half-line. In this paper, we
assume that the solution $u(x,t)$ and $v(x,t)$ of CDNLS equations are
exists, and we show that it can be expressed in terms of the unique
solution of a matrix Riemann-Hilbert problem formulated in the plane
of the complex spectral parameter $\lambda$.

\section{Introduction}
\quad\;\;Since the 1960s, the inverse scattering method with the initial
value problem give mathematical physicists great power to find exact solution
of the nonlinear partial differential equations. This includes continuity and
discrete partial differential equations (usually we call soliton equations
because they have soliton solutions). Thanks to four mathematicians Gardner,
Greene, Kruskal and Miura. They initially studied KdV equation exact solution
(1967,1974) with the initial value problem. Because this method can have the
power of infinite life for the whole family of equations, and has been applied
to many scientific and technological fields including geophysical prospecting,
super symmetric quantum mechanics and so on. But for the initial boundary value(IBV)
problem of the soliton equation, how to find their exact solution? This is a
very big challenge problem. Fortunately, in general as long as the equation is
integrable, these problems can be solved. Here integrable are equations which have Lax pairs.

In 1997, Fokas used inverse scattering transform(IST) thought to construct a
new unified method, we call this method as Fokas method. He analyzed the IBV
problems for linear and nonlinear integrable PDEs\cite{Fokas1997,Fokas2002,Fokas2008}.
In the past 20 years, the unified method has been used to analyse boundary value
problems for several of the most important integrable equations with $2\times2$
matrix Lax pairs, such as the Korteweg-deVries(KdV) equation, the nonlinear
Schr\"{o}dinger(NLS) equation, the sine-Gordon(sG) equation, the derivative NLS(DNLS) equation and the complex
Sharma-Tasso-Olver(CSTO) equation [4-8], etc.
Just like the IST on the line, the unified method provides an expression for
the solution of an IBV problem in terms of the solution of a Riemann-Hilbert
problem. In particular, by analyzing the asymptotic behaviour of the solution
based on this Riemann-Hilbert problem and by employing the nonlinear version
of the steepest descent method introduced by Deift and Zhou \cite{Deift1993-1} in 1993.
In this way, the long time asymptotics for the solutions of decay initial value problem of NLS equation and the MKdV equation
were analyzed respectively by Deift and Zhou \cite{Deift1993-1,Deift1993-2}. The DNLS and other
integrable equations have been rigorously established [11-14]. For the asymptotics of the solution of IBV problem and step-like
initial value problem for the DNLS also have been considered \cite{Arruda2017,Xujian2013}.

In 2012, Lenells\cite{Lenells2012} applied the unified transform method to analyse
IBV problems for integrable evolution equations whose Lax pairs involving $3\times3$
matrices, and following this method, the IBV problems for the Degasperis-Procesi
equation to be studied in\cite{Lenells2013}. After that, most important integrable
equations IBV problems for integrable evolution equations with higher order Lax
pairs to be studied [19-28]. we also have a good time to study partial differential
equations with IBV problem on the basis of these giants.
In this paper, we would like to analyse the IBV problem of the following coupled DNLS
equation\cite{Morr1979,His1995,Zhang2009}:
\eqa \left\{ \begin{array}{l}
iu_t+u_{xx}+i\gamma[(|u|^2+|v|^2)u]_x=0,\\
iv_t+v_{xx}+i\gamma[(|u|^2+|v|^2)v]_x=0.
\end{array} \right. \label{slisp}\eeqa
on the half-line domain $\Omega=\{0<x<\infty,0<t<T\}$.
Throughout this paper, we consider the following IBV problems for the CDNLS equations
\eqa\begin{array}{l}
$Initial values$: u_0(x)=u(x,t=0),\; v_0(x)=v(x,t=0);\\
$Dirichlet boundary values$: g_0(t)=u(x=0,t),\; h_0(t)=v(x=0,t);\\
$Neumann boundary values$: g_1(t)=u_x(x=0,t),\; h_1(t)=v_x(x=0,t).
\end{array}\label{slisp}\eeqa
where $u_0(x)$ and $v_0(x)$ lie in the Schwartz space.

This paper is organized as follows. In the next section, we define two sets of
eigenfunctions $\mu_j(j=1,2,3)$ and $M_n(n=1,2,3,4)$ of Lax pair for spectral
analysis. In addition, we also get some spectral functions satisfying the
so-called global relation in this part. In the last section, we show that
$u(x,t),v(x,t)$ can be expressed in terms of the unique solution of a matrix Riemann-Hilbert problem.

\section{ The spectral analysis}

\quad\;\;Consider the Lax pair of equations (1.1) as follows\cite{Li2010}
 \eqa \left\{ \begin{array}{l}
\psi_x=U\psi=(-\frac{1}{\gamma}i\lambda^2\Lambda+\lambda U_1)\psi,\\
\psi_t=V\psi=(\frac{2}{\gamma^2}i\lambda^4\Lambda-\frac{2}{\gamma}\lambda^3U_1+i\lambda^2U_2-\gamma\lambda U_3)\psi,
\end{array} \right. \label{slisp}\eeqa
where
\eqa\begin{array}{l}
\Lambda=\left(\begin{array}{ccc}
-1&0&0\\
0&1&0\\
0&0&1\end{array} \right),
U_1=\left(\begin{array}{ccc}
0& u& v\\
\bar u&0&0\\
\bar v&0&0\end{array} \right),
U_2=\left(\begin{array}{ccc}
-(|u|^2+|v|^2)& 0& 0\\
0&|u|^2&\bar uv\\
0&u\bar v&|v|^2\end{array} \right),\\
U_3=\left(\begin{array}{ccc}
0& u(|u|^2+|v|^2)-iu_x& v(|u|^2+|v|^2)-iv_x\\
\bar u(|u|^2+|v|^2)+i\bar u_x&0&0\\
\bar v(|u|^2+|v|^2)+i\bar v_x&0&0\end{array} \right).
\end{array}\label{slisp}\eeqa
where the overbar represents the complex conjugation (similarly hereinafter), $\lambda$ is a spectral
parameter, and $\psi(x,t,\lambda)$ is a vector or a matrix function. Throughout this paper, we set $\gamma=1$ for the convenient of the analysis.

\subsection{The closed one-form}

\quad\;\;We are not difficult to find that Eq.(2.1) is equivalent to
\eqa \left\{ \begin{array}{l}
\psi_x+i\lambda^2\Lambda\psi=V_1\psi,\\
\psi_t-2i\lambda^4\Lambda\psi=V_2\psi,
\end{array} \right. \label{slisp}\eeqa
where
 \beq V_1=\lambda U_1,\quad V_2=-2\lambda^3U_1+i\lambda^2U_2-\lambda U_3.\label{slisp}\eeq

We assume that $u(x,t),v(x,t)$ is a sufficiently smooth function in the half-line region $\Omega=\{0<x<\infty,0<t<T\}$, and decays sufficiently when $x\rightarrow\infty.$ Introducing a new function $\mu(x,t,\lambda)$ by
\beq\psi=\mu e^{-i\lambda^2\Lambda x+2i\lambda^4\Lambda t},\label{slisp}\eeq
then the Lax pair Eq.(2.3) becomes
\eqa \left\{ \begin{array}{l}
\mu_x+i\lambda^2[\Lambda,\mu]=V_1\mu,\\
\mu_t-2i\lambda^4[\Lambda,\mu]=V_2\mu,
\end{array} \right. \label{slisp}\eeqa
and Eq.(2.6) can be written to the differential form
\beq d(e^{i\lambda^2\hat\Lambda x-2i\lambda^4\hat\Lambda t}\mu)=W(x,t,\lambda), \label{slisp}\eeq
where $W(x,t,\lambda)$ defined by
\beq W(x,t,\lambda)=e^{(i\lambda^2 x-2i\lambda^4t)\hat\Lambda}(V_1dx+V_2dt)\mu, \label{slisp}\eeq
and $\hat\Lambda$ represents a matrix operator acting on $3\times3$ matrix B by $\hat\Lambda B=[\Lambda,B]$.

\subsection{ The eigenfunction}

\quad\;\;Based on the Volterra integral equation, there are three eigenfunctions $\mu_j(x,t,\lambda)(j=1,2,3)$ of Eq.(2.6) defined as
\beq \mu_j(x,t,\lambda)=\mathbb{I}+\int_{\gamma_j}e^{(-i\lambda^2 x+2i\lambda^4t)\hat\Lambda}W_j(\xi,\tau,\lambda),\quad j=1,2,3,\label{slisp}\eeq
where $W_j$ is determined Eq.(2.8), it is only used $\mu_j$ in place of $\mu$, and the contours $ \gamma_j(j=1,2,3)$ are shown in figure 1.

\begin{figure}
\centering
\includegraphics[width=3.8in,height=1.1in]{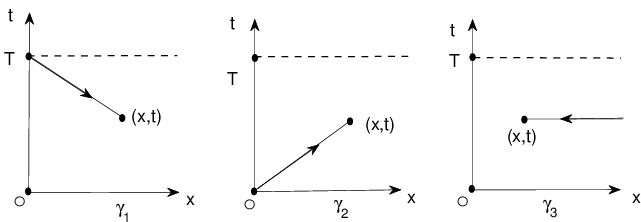}
\caption{The three contours $\gamma_1,\gamma_2,\gamma_3$ in the $(x,t)$-domaint}
\label{fig:graph}
\end{figure}

The first, second, and third columns of the matrix equation (2.9) contain the following exponential term
\eqa\begin{array}{l}
\mu_j^{(1)}:\quad e^{-2i\lambda^2(x-\xi)+4i\lambda^4(t-\tau)}, e^{-2i\lambda^2(x-\xi)+4i\lambda^4(t-\tau)};\\
\mu_j^{(2)}:\quad e^{2i\lambda^2(x-\xi)-4i\lambda^4(t-\tau)};\\
\mu_j^{(3)}:\quad e^{2i\lambda^2(x-\xi)-4i\lambda^4(t-\tau)}.
\end{array}\label{slisp}\eeqa
At the same time, the following inequalities hold true on the contours
\eqa \begin{array}{l}
\gamma_1:\quad x-\xi\geq 0,\quad t-\tau\leq 0;\\
\gamma_2:\quad x-\xi\geq 0,\quad t-\tau\geq 0;\\
\gamma_3:\quad x-\xi\leq 0.\\
\end{array}\label{slisp}\eeqa
Thus, we can show that the eigenfunctions $\mu_j(x,t,\lambda)(j=1,2,3)$ are bounded and analytic for $\lambda\in\mathbb{C}$ such that $\lambda$ belongs to
\eqa\begin{array}{l}
 \mu_1 $ is bounded and analytic for $ \lambda\in(D_4,D_1,D_1),\\
 \mu_2 $ is bounded and analytic for $ \lambda\in(D_3,D_2,D_2),\\
 \mu_3 $ is bounded and analytic for $ \lambda\in(D_1\cup D_2,D_3\cup D_4,D_3\cup D_4),\\
\end{array}\label{slisp}\eeqa
where $D_n(n=1,2,3,4)$ denote four open, pairwisely disjoint subsets of the Riemann $\lambda$-plane shown in figure 2.

\begin{figure}
\centering
\includegraphics[width=3.0in,height=2.0in]{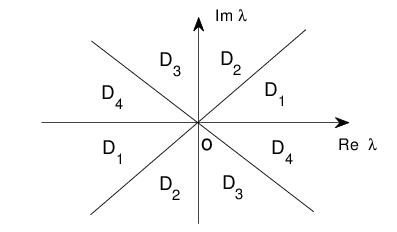}
\caption{The sets $D_j,j=1,2,3,4$, which decompose the complex $\lambda-$plane}
\label{fig:graph}
\end{figure}

And these sets $D_n(n=1,2,3,4)$ have the following properties:
\eqa\begin{array}{l}
D_1=\{\lambda\in\mathbb{C}|Rel_1<Rel_2=Rel_3,\quad Rez_1>Rez_2=Rez_3\},\\
D_2=\{\lambda\in\mathbb{C}|Rel_1<Rel_2=Rel_3,\quad Rez_1<Rez_2=Rez_3\},\\
D_3=\{\lambda\in\mathbb{C}|Rel_1>Rel_2=Rel_3,\quad Rez_1>Rez_2=Rez_3\},\\
D_4=\{\lambda\in\mathbb{C}|Rel_1>Rel_2=Rel_3,\quad Rez_1<Rez_2=Rez_3\},
\end{array}\label{slisp}\eeqa
where $l_i(\lambda)$ and $z_i(\lambda)$ are the diagonal elements of the matrix $-i\lambda^2\Lambda$ and $2i\lambda^4\Lambda$.

Specially, we can show that $\mu_1(0,t,\lambda)$ is bounded and analytic for $\lambda\in(D_2\cup D_4,D_1\cup D_3,D_1\cup D_3)$, and $\mu_2(0,t,\lambda)$ is bounded and analytic for $\lambda\in(D_1\cup D_3,D_2\cup D_4,D_2\cup D_4)$.

For each $n=1,2,3,4$, based on the following integral equation, the solution $M_n(x,t,\lambda)$ of Eq.(2.6) can be defined as
\beq (M_n(x,t,\lambda))_{ij}=\delta_{ij}+\int_{\gamma_{ij}^n}(e^{(-i\lambda^2 x+2i\lambda^4t)\hat\sigma}W_n(\xi,\tau,\lambda))_{ij},
\quad i,j=1,2,3,\label{slisp}\eeq
where $W_n(x,t,\lambda)$ is determined by Eq.(2.8), it is only used $M_n$ in place of $\mu$,
and the contours $ \gamma_{ij}^n (n=1,2,3,4;i,j=1,2,3)$ are defined as
\beq \gamma_{ij}^n=\left\{\begin{array}{l}
\gamma_1\quad if\quad Rel_i(\lambda)<Rel_j(\lambda) \quad and \quad Rez_i(\lambda)\geq Rez_j(\lambda),\\
\gamma_2\quad if\quad Rel_i(\lambda)<Rel_j(\lambda) \quad and \quad Rez_i(\lambda)<Rez_j(\lambda), for\lambda\in D_n\\
\gamma_3\quad if \quad Rel_i(\lambda)\geq Rel_j(\lambda).
\end{array} \right. \label{slisp} \eeq
Based on the definition of $\gamma^n$, we have
\eqa \begin{array}{l}
\gamma^1=\left(\begin{array}{ccc}
\gamma_3&\gamma_1&\gamma_1\\
\gamma_3&\gamma_3&\gamma_3\\
\gamma_3&\gamma_3&\gamma_3\end{array} \right),\quad
\gamma^2=\left(\begin{array}{ccc}
\gamma_3&\gamma_2&\gamma_2\\
\gamma_3&\gamma_3&\gamma_3\\
\gamma_3&\gamma_3&\gamma_3\end{array} \right),\\
\gamma^3=\left(\begin{array}{ccc}
\gamma_3&\gamma_3&\gamma_3\\
\gamma_2&\gamma_3&\gamma_3\\
\gamma_2&\gamma_3&\gamma_3\end{array} \right),\quad
\gamma^4=\left(\begin{array}{ccc}
\gamma_3&\gamma_3&\gamma_3\\
\gamma_1&\gamma_3&\gamma_3\\
\gamma_1&\gamma_3&\gamma_3\end{array} \right).
\end{array}\label{slisp} \eeqa

Next, the following proposition guarantees that the previous definition of $M_n$ has properties, namely, $M_n$ can be represented as a Rimann-Hilbert problem.

\begin{prop} For each $n=1,2,3,4$ and $\lambda\in D_n$, the function $M_n(x,t,\lambda)$ is defined well by Eq.(2.14).
And for any identified point $(x,t)$, $M_n$ is bounded and analytical as a function of $\lambda\in D_n$ away from a possible discrete set of singularities $\{\lambda_j\}$ at which the Fredholm determinant vanishes. Moreover, $M_n$ admits a bounded and continuous extension to $\bar D_n$ and
\beq M_n(x,t,\lambda)=\mathbb{I}+\mathcal{O}(\frac{1}{\lambda}).\label{slisp}\eeq\end{prop}

Proof: The associated bounded and analytical properties have been established in Appendix B in \cite{Lenells2012}. Substituting the following expansion
\beq M=M_0+\frac{M^{(1)}}{\lambda}+\frac{M^{(2)}}{\lambda^2}+\cdots\quad\quad\lambda\rightarrow\infty,\nn\eeq
into the Lax pair Eq.(2.6) and comparing the coefficients of the same order of $\lambda$, we can obtain Eq.(2.17).

\subsection{ The jump matrix}

\quad\;\;Define the matrix-value functions as follows
\beq S_n(\lambda)=M_n(0,0,\lambda),\quad \lambda\in D_n,n=1,2,3,4.\label{slisp}\eeq
Let $M$ be a sectionally analytical continuous function in Riemann $\lambda$-sphere which equals $M_n$ for
$\lambda\in D_n$. Then $M$ satisfies the following jump conditions:
\beq M_n(\lambda)=M_mJ_{m,n},\quad \lambda\in \bar D_n\cap \bar D_m,\quad n,m=1,2,3,4;n\neq m,\label{slisp}\eeq
where
\beq J_{m,n}=e^{(-i\lambda^2 x+2i\lambda^4t)\hat\Lambda}(S_m^{-1}S_n).\label{slisp}\eeq

\subsection{ The adjugated eigenfunction}

\quad\;\;To obtain the analyticity and boundedness properties of the minors of the matrices $\mu_j(x,t,\lambda)(j=1,2,3)$. We need consider the cofactor matrix $B^A$ of a $3\times3$ matrix $B$ is defined by
$$ B^A=\left(\begin{array}{ccc}
m_{11}(B)&-m_{12}(B)&m_{13}(B)\\
-m_{21}(B)&m_{22}(B)&-m_{23}(B)\\
m_{31}(B)&-m_{32}(B)&m_{33}(B) \end{array}
\right),$$
where $m_{ij}(B)$ denote the $(ij)$th minor of $B$.

From Eq.(2.6) we find that the adjugated eigenfunction $\mu^A$ have the following Lax pair equations:
\eqa \left\{\begin{array}{l}
\mu_x^A-i\lambda^2[\Lambda,\mu^A]=-V_1^T\mu^A,\\
\mu_t^A+2i\lambda^4[\Lambda,\mu^A]=-V_2^T\mu^A,\\
\end{array}\right. \label{slisp}\eeqa
where the superscript $T$ denotes a matrix transpose. Then the eigenfunctions $\mu_j(j=1,2,3)$ satisfy the following integral equations
\beq \mu_j^A(x,t,\lambda)=\mathbb{I}-\int_{\gamma_j}e^{(i\lambda^2(x-\xi)-2i\lambda^4(t-\tau))\hat\Lambda}(V_1^Tdx+V_2^Tdt),\quad j=1,2,3.\label{slisp}\eeq
Thus, we can obtain the adjugated eigenfunction satisfies the following analyticity and boundedness properties:
\eqa\begin{array}{l}
 \mu_1^A $ is bounded and analytic for $ \lambda\in(D_1,D_4,D_4),\\
 \mu_2^A $ is bounded and analytic for $ \lambda\in(D_2,D_3,D_3),\\
 \mu_3^A $ is bounded and analytic for $ \lambda\in(D_3\cup D_4,D_1\cup D_2,D_1\cup D_2).
\end{array}\label{slisp}\eeqa

Specially, we can show that $\mu_1^A(0,t,\lambda)$ is bounded and analytic for $\lambda\in(D_1\cup D_3,D_2\cup D_4,D_2\cup D_4)$, and $\mu_2^A(0,t,\lambda)$ is bounded and analytic for $\lambda\in(D_2\cup D_4,D_1\cup D_3,D_1\cup D_3)$.

\subsection{ Symmetry}

\quad\;\; We can show that the eigenfunctions $\mu_j(x,t,\lambda)$ have an important symmetry by the following Lemma.
\begin{lem} The eigenfunction $\psi(x,t,\lambda)$ of the Lax pair Eq.(2.1) have the following symmetry
$$\psi^{-1}(x,t,\lambda)=A \overline{\psi(x,t,\bar\lambda)}^TA,$$
with
$$
A=\left(\begin{array}{ccc}
1&0&0\\
0&-\varepsilon&0\\
0&0&-\varepsilon\end{array}
\right),\quad and \quad\varepsilon^2=1.$$
where the superscript $T$ denotes a matrix transpose.\end{lem}

Proof: Analogous to the proof provided in \cite{Lenells2012}. We omit the proof.
\begin{rem} From Lemma 2.2, we can show that the eigenfunctions $\mu_j(x,t,\lambda)$ of Lax pair Eq.(2.6) have the same symmetry.\end{rem}

\subsection{ The jump matrix computations}

\quad\;\;We also define the $3\times3$ matrix value spectral function $s(\lambda)$ and $S(\lambda)$ as follows
\eqa \left\{ \begin{array}{l}
\mu_3(x,t,\lambda)=\mu_2(x,t,\lambda)e^{(-i\lambda^2 x+2i\lambda^4t)\hat\Lambda} s(\lambda),\\
\mu_1(x,t,\lambda)=\mu_2(x,t,\lambda)e^{(-i\lambda^2 x+2i\lambda^4t)\hat\Lambda} S(\lambda),
\end{array}\right. \label{slisp}\eeqa
by $\mu_2(0,0,\lambda)=\mathbb{I}$, and from Eq.(2.24) we can obtain
\beq s(\lambda)=\mu_3(0,0,\lambda),\quad\quad S(\lambda)=\mu_1(0,0,\lambda).\label{slisp}\eeq

From the properties of $\mu_j$ and $\mu_j^A(j=1,2,3)$, we can drive that $s(\lambda),S(\lambda),s^A(\lambda)$
and $S^A(\lambda)$ have the following bounded and analytic properties
\eqa\begin{array}{l} s(\lambda) $ is bounded for $ \lambda\in(D_1\cup D_2,D_3\cup D_4,D_3\cup D_4),\\
 S(\lambda)$ is bounded for $ \lambda\in(D_2\cup D_4,D_1\cup D_3,D_1\cup D_3),\\
 s^A(\lambda)$ is bounded for $ \lambda\in(D_3\cup D_4,D_1\cup D_2,D_1\cup D_2),\\
 S^A(\lambda)$ is bounded for $ \lambda\in(D_1\cup D_3,D_2\cup D_4,D_2\cup D_4).
\end{array}\label{slisp}\eeqa
Moreover
\beq M_n(x,t,\lambda)=\mu_2(x,t,\lambda)e^{(-i\lambda^2 x+2i\lambda^4t)\hat\Lambda} S_n(\lambda),\quad \lambda\in D_n.\label{slisp}\eeq
\begin{prop}
The $S_n$ can be expressed with $s(\lambda)$ and $S(\lambda)$  elements as follows
\eqa \begin{array}{l}  S_1=\left(\begin{array}{ccc}
s_{11}&\frac{m_{33}(s)M_{21}(S)-m_{23}(s)M_{31}(S)}{(s^TS^A)_{11}}&\frac{m_{32}(s)M_{21}(S)-m_{22}(s)M_{31}(S)}{(s^TS^A)_{11}}\\
s_{21}&\frac{m_{33}(s)M_{11}(S)-m_{13}(s)M_{31}(S)}{(s^TS^A)_{11}}&\frac{m_{32}(s)M_{11}(S)-m_{12}(s)M_{31}(S)}{(s^TS^A)_{11}}\\
s_{31}&\frac{m_{23}(s)M_{11}(S)-m_{13}(s)M_{21}(S)}{(s^TS^A)_{11}}&\frac{m_{22}(s)M_{11}(S)-m_{12}(s)M_{21}(S)}{(s^TS^A)_{11}}
\end{array} \right),\\
S_2=\left(\begin{array}{ccc}
s_{11}&0&0\\
s_{21}&\frac{m_{33}(s)}{s_{11}}&\frac{m_{32}(s)}{s_{11}}\\
s_{31}&\frac{m_{23}(s)}{s_{11}}&\frac{m_{22}(s)}{s_{11}}
\end{array} \right),\quad
S_3=\left(\begin{array}{ccc}
\frac{1}{m_{11}(s)}&s_{12}&s_{13}\\
0&s_{22}&s_{23}\\
0&s_{32}&s_{33}
\end{array} \right),\\
S_4=\left(\begin{array}{ccc}
\frac{S_{11}}{(S^Ts^A)_{11}}&s_{12}&s_{13}\\
\frac{S_{21}}{(S^Ts^A)_{11}}&s_{22}&s_{23}\\
\frac{S_{31}}{(S^Ts^A)_{11}}&s_{32}&s_{33}
\end{array} \right).\end{array}\label{slisp}\eeqa
\end{prop}
where $(S^Ts^A)_{11}$ and $(s^TS^A)_{11}$ are defined as follows
\eqa\begin{array}{l}
(S^Ts^A)_{11}=S_{11}m_{11}(s)-S_{21}m_{21}(s)+S_{31}m_{31}(s),\\
(s^TS^A)_{11}=s_{11}m_{11}(S)-s_{21}m_{21}(S)+s_{31}m_{31}(S).
\end{array}\label{slisp}\eeqa

Proof: We set that $\gamma_3^{X_0}$ is a contour when $(X_0,0)\rightarrow (x,t)$ in the $(x,t)$-plane, here $X_0$ is a constant and $X_0>0$, for $j=3$, we introduce $\mu_3(x,t,\lambda;X_0)$ as the solution of Eq.(2.9) with the contour $\gamma_3$ replaced by $\gamma_3^{X_0}$. Similarly, we define $M_n(x,t,\lambda;X_0)$ as the solution of Eq.(2.14) with $\gamma_3$ replaced by $\gamma_3^{X_0}$. then, by simple calculation, we can use $S(\lambda)$ and $s(\lambda;X_0)=\mu_3(0,0,\lambda;X_0)$ to derive the expression of $S_n(\lambda,X_0)=M_n(0,0,\lambda;X_0)$ and the Eq.(2.28) will be obtained by taking the limit $X_0\rightarrow\infty$.

Firstly, we have the following relations:
\eqa
M_n(x,t,\lambda;X_0)=\mu_1(x,t,\lambda)e^{(-i\lambda^2 x+2i\lambda^4t)\hat\Lambda} R_n(\lambda;X_0),\\
M_n(x,t,\lambda;X_0)=\mu_2(x,t,\lambda)e^{(-i\lambda^2 x+2i\lambda^4t)\hat\Lambda} S_n(\lambda;X_0),\\
M_n(x,t,\lambda;X_0)=\mu_3(x,t,\lambda)e^{(-i\lambda^2 x+2i\lambda^4t)\hat\Lambda} T_n(\lambda;X_0).
\label{slisp}\eeqa

Secondly, we can get the definition of $R_n(\lambda;X_0)$ and $T_n(\lambda;X_0)$ as follows
\eqa
R_n(\lambda;X_0)=e^{-2i\lambda^4T\hat\Lambda} M_n(0,T,\lambda;X_0),\\
T_n(\lambda;X_0)=e^{i\lambda^2 X_0\hat\Lambda} M_n(X_0,0,\lambda;X_0),
\label{slisp}\eeqa
then equations(2.29),(2.30) and (2.31) mean that
\eqa
s(\lambda;X_0)=S_n(\lambda;X_0)T_n^{-1}(\lambda;X_0),\;S(\lambda;X_0)=S_n(\lambda;X_0)R_n^{-1}(\lambda;X_0).
\label{slisp}\eeqa

These equations constitute the matrix decomposition problem of $\{s,S\}$ by use $\{R_n,S_n,T_n\}$. In fact, by the definition of the integral equation (2.14) and $\{R_n,S_n,T_n\}$, we obtain
\beq \left\{ \begin{array}{l}
(R_n(\lambda;X_0))_{ij}=0\quad if \quad \gamma_{ij}^n=\gamma_1,\\
(S_n(\lambda;X_0))_{ij}=0\quad if \quad \gamma_{ij}^n=\gamma_2,\\
(T_n(\lambda;X_0))_{ij}=\delta_{ij}\quad if \quad \gamma_{ij}^n=\gamma_3.
\end{array}\right. \label{slisp}\eeq

Thus equations (2.35) is the 18 scalar equations with 18 unknowns. The exact solution of these system can be obtained by solving the algebraic system,
 in this way, we can get a similar $\{S_n(\lambda),s(\lambda)\}$ as in Eq.(2.28) which just that $\{S_n(\lambda),s(\lambda)\}$ replaces by $\{S_n(\lambda;X_0),s(\lambda;X_0)\}$ in Eq.(2.28).

 Finally, taking $X_0\rightarrow\infty$ in this equation, we obtain the Eq.(2.28).

\subsection{ The residue conditions}

\quad\;\;Because $\mu_2$ is an entire function, and from Eq.(2.27) we know that $M$ only produces singularities in $S_n$ where there are singular points, from the exact expression Eq.(2.28), we know that $M$ may be singular as follows

(1)\quad$[M_1]_2$ and $[M_1]_3$ could have poles in $D_1$ at the zeros of $(s^TS^A)_{11}(\lambda)$,

(2)\quad$[M_2]_2$ and $[M_2]_3$ could have poles in $D_2$ at the zeros of $s_{11}(\lambda)$,

(3)\quad$[M_3]_1$ could have poles in $D_3$ at the zeros of $m_{11}(s)(\lambda)$,

(4)\quad$[M_4]_1$ could have poles in $D_4$ at the zeros of $(S^Ts^A)_{11}(\lambda)$.

We use $\lambda_j(j=1,2\cdots N)$ denote the possible zero point above, and assume that these zeros satisfy the following assumptions
\begin{ass}We assume that\end{ass}

(1)$(s^TS^A)_{11}(\lambda)$ has $n_0$ possible simple zeros in $D_1$ denoted by $\lambda_j,j=1,2\cdots n_0$,

(2)$s_{11}(\lambda)$ has $n_1-n_0$ possible simple zeros in $D_2$ denoted by $\lambda_j,j=n_0+1,n_0+2\cdots n_1$,

(3)$m_{11}(s)(\lambda)$ has $n_2-n_1$ possible simple zeros in $D_3$ denoted by $\lambda_j,j=n_1+1,n_1+2\cdots n_2$,

(4)$(S^Ts^A)_{11}(\lambda)$ has $N-n_2$ possible simple zeros in $D_4$ denoted by $\lambda_j,j=n_2+2,n_2+2\cdots N$,

And these zeros are each different, moreover assuming that there is no zero on the boundary of $D_n(n=1,2,3,4)$.

\begin{prop}
Let $M_n(n=1,2,3,4)$ be the eigenfunctions defined by (2.14) and assume that the set $\lambda_j(j=1,2\cdots N)$  of singularities are as the above assumption. Then the following residue conditions hold true:

\beq \begin{array}{l} Res_{\lambda=\lambda_j}[M]_2=\frac{m_{33}(s)(\lambda_j)M_{11}(S)(\lambda_j)-m_{13}(s)(\lambda_j)M_{31}(S)(\lambda_j)}{\dot{(s^TS^A)_{11}(\lambda_j)}s_{21}(\lambda_j)}e^{\theta_{13}(\lambda_j)}[M(\lambda_j)]_1
,\\ \quad\quad\quad\quad\quad\quad\quad\quad\quad\quad\quad\quad\quad\quad\quad\quad\quad\quad\quad\quad
 1\leq j\leq n_0;\lambda_j\in D_1.\end{array}\label{slisp}\eeq
\beq \begin{array}{l} Res_{\lambda=\lambda_j}[M]_3=\frac{m_{32}(s)(\lambda_j)M_{11}(S)(\lambda_j)-m_{12}(s)(\lambda_j)M_{31}(S)(\lambda_j)}{\dot{(s^TS^A)_{11}(\lambda_j)}s_{21}(\lambda_j)}e^{\theta_{13}(\lambda_j)}[M(\lambda_j)]_1
,\\ \quad\quad\quad\quad\quad\quad\quad\quad\quad\quad\quad\quad\quad\quad\quad\quad\quad\quad\quad\quad
1\leq j\leq n_0;\lambda_j\in D_1.\end{array}\label{slisp}\eeq
\beq \begin{array}{l}
Res_{\lambda=\lambda_j}[M]_2=\frac{m_{33}(s)(\lambda_j)}{\dot{s_{11}(\lambda_j)}s_{21}(\lambda_j)}e^{\theta_{13}(\lambda_j)}[M(\lambda_j)]_1
,\quad n_0+1\leq j\leq n_1;\lambda_j\in D_2.\end{array}\label{slisp}\eeq
\beq \begin{array}{l}
Res_{\lambda=\lambda_j}[M]_3=\frac{m_{32}(s)(\lambda_j)}{\dot{s_{11}(\lambda_j)}s_{21}(\lambda_j)}e^{\theta_{13}(\lambda_j)}[M(\lambda_j)]_1
,\quad n_0+1\leq j\leq n_1;\lambda_j\in D_2.\end{array}\label{slisp}\eeq
\beq \begin{array}{l} Res_{\lambda=\lambda_j}[M]_1=\frac{s_{33}(\lambda_j)[M_(\lambda_j)]_2-s_{32}(\lambda_j)[M_(\lambda_j)]_3}{\dot{m_{11}(s)(\lambda_j)}m_{21}(s)(\lambda_j)}e^{\theta_{31}(\lambda_j)}
, n_1+1\leq j\leq n_2;\lambda_j\in D_3.\end{array}\label{slisp}\eeq
\beq \begin{array}{l} Res_{\lambda=\lambda_j}[M]_1=\frac{s_{33}(\lambda_j)S_{21}(\lambda_j)-s_{23}(\lambda_j)S_{31}(\lambda_j)}{\dot{(S^Ts^A)_{11}(\lambda_j)}m_{11}(s)(\lambda_j)}e^{\theta_{31}(\lambda_j)}[M(\lambda_j)]_2
\\ \quad\quad\quad\quad\quad\quad
+\frac{s_{22}(\lambda_j)S_{31}(\lambda_j)-s_{32}(\lambda_j)S_{21}(\lambda_j)}{\dot{(S^Ts^A)_{11}(\lambda_j)}m_{11}(s)(\lambda_j)}e^{\theta_{31}(\lambda_j)}[M(\lambda_j)]_3
,\\ \quad\quad\quad\quad\quad\quad\quad\quad\quad\quad\quad\quad\quad\quad\quad\quad\quad\quad\quad n_2+1\leq j\leq N;\lambda_j\in D_4.\end{array}\label{slisp}\eeq
where $\dot{f}=\frac{df}{d\lambda}$ and $\theta_{ij}$ defined by
\beq \theta_{ij}(x,t,\lambda)=(l_i-l_j)x-(z_i-z_j)t, \quad i,j=1,2,3;\label{slisp}\eeq
thus
\beq \theta_{ij}=0,\quad i,j=2,3;\quad\quad \theta_{12}=\theta_{13}=-\theta_{21}=-\theta_{31}=2i\lambda^2 x+4i\lambda^4t.\nn\eeq
\end{prop}

Proof: We will only prove (2.41), (2.42) and the other conditions follow by similar arguments. The equation (2.27) mean that
\eqa
M_3=\mu_2e^{(-i\lambda^2 x+2i\lambda^4t)\hat\Lambda} S_3,\\
M_4=\mu_2e^{(-i\lambda^2 x+2i\lambda^4t)\hat\Lambda} S_4.
\label{slisp}\eeqa

In view of the expressions for $S_3$ given in (2.28), the three columns of Eq.(2.44) read
\beq [M_3]_1=\frac{1}{m_{11}(s)}[\mu_2]_1,\label{slisp}\eeq
\beq [M_3]_2=[\mu_2]_1s_{12}e^{\theta_{13}}+[\mu_2]_2s_{22}+[\mu_2]_3s_{32},\label{slisp}\eeq
\beq [M_3]_3=[\mu_2]_1s_{13}e^{\theta_{13}}+[\mu_2]_2s_{23}+[\mu_2]_3s_{33}.\label{slisp}\eeq

And in view of the expressions for $S_4$ given in (2.28), the three columns of Eq.(2.45) read
\beq [M_4]_1=\frac{S_{11}}{(S^Ts^A)_{11}}[\mu_2]_1+\frac{S_{21}}{(S^Ts^A)_{11}}[\mu_2]_2e^{\theta_{31}}
+\frac{S_{31}}{(S^Ts^A)_{11}}[\mu_2]_3e^{\theta_{31}},\label{slisp}\eeq
\beq [M_4]_2=[\mu_2]_1s_{12}e^{\theta_{13}}+[\mu_2]_2s_{22}+[\mu_2]_3s_{32},\label{slisp}\eeq
\beq [M_4]_3=[\mu_2]_1s_{13}e^{\theta_{13}}+[\mu_2]_2s_{23}+[\mu_2]_3s_{33}.\label{slisp}\eeq

We suppose that $\lambda_j\in D_3$ is a simple zero of $m_{11}(s)(\lambda)$. Solving Eq.(2.47) and  Eq.(2.48) for $[\mu_2]_1$ substituting
the result into Eq.(2.46), we find
\beq [M_3]_1=\frac{s_{33}[M_3]_2-s_{32}[M_3]_3}{m_{11}(s)m_{21}(s)}e^{\theta_{31}}-\frac{1}{m_{21}(s)}e^{\theta_{31}}[\mu_2]_2.\label{slisp}\eeq
Taking the residue of this equation at $\lambda_j$, we find condition Eq.(2.41) in the case when $\lambda_j\in D_3$.

In the same way, we assume that $\lambda_j\in D_4$ is a simple zero of $(S^Ts^A)_{11}(\lambda)$. Solving Eq.(2.50) and  Eq.(2.51) for $[\mu_2]_2$ and $[\mu_2]_3$ substituting the result into Eq.(2.49), we find
\beq [M_4]_1=\frac{s_{33}S_{21}-s_{23}S_{31}}{(s^TS^A)_{11}m_{11}(s)}e^{\theta_{31}}[M_4]_2+\frac{s_{33}S_{21}-s_{23}S_{31}}{(s^TS^A)_{11}m_{11}(s)}e^{\theta_{31}}[M_4]_3
+\frac{1}{m_{11}(s)}[\mu_2]_1.\label{slisp}\eeq
Taking the residue of this equation at $\lambda_j$, we find condition Eq.(2.42) in the case when $\lambda_j\in D_4$.

\subsection{ The global relation }

\quad\;\;The spectral functions $S(\lambda)$ and $s(\lambda)$ are not independent which is of important relationship each other. In fact, from Eq.(2.24), we find
\beq \mu_3(x,t,\lambda)=\mu_1(x,t,\lambda)e^{(-i\lambda^2 x+2i\lambda^4t)\hat\Lambda} S^{-1}(\lambda)s(\lambda),
\lambda\in(D_1\cup D_2,D_3\cup D_4,D_3\cup D_4),\label{slisp}\eeq
as $\mu_1(0,t,\lambda)=\mathbb{I}$, when $(x,t)=(0,T)$, We can evaluate the following relationship which is the global relation as follows
\beq S^{-1}(\lambda)s(\lambda)=e^{-2i\lambda^4T\hat\Lambda}c(T,\lambda),\quad
\lambda\in(D_1\cup D_2,D_3\cup D_4,D_3\cup D_4),
\label{slisp}\eeq
where  $c(T,\lambda)=\mu_3(0,t,\lambda)$.

\section{The Riemann-Hilbert problem }

\quad\;\;In section 2, we define the sectionally analytical function $M(x,t,\lambda)$ that its satisfies a Riemann-Hilbert problem which can be formulated in terms of the initial and boundary values of $\{u(x,t),v(x,t)\}$. For all $(x,t)$, the solution of Eq.(1.1) can be recovered by solving this Riemann-Hilbert problem. So we can establish the following theorem.

\begin{thm}Suppose that $\{u(x,t),v(x,t)\}$ are solution of Eq.(1.1) in the half-line domain $\Omega$, and it is sufficient smoothness and decays when $x\rightarrow\infty$. Then the $\{u(x,t),v(x,t)\}$ can be reconstructed from the initial values $\{u_0(x),v_0(x)\}$ and boundary values $\{g_0(t),h_0(t),g_1(t),h_1(t)\}$ defined as follows
\eqa\begin{array}{l}
u_0(x)=u(x,0),\quad v_0(x)=v(x,0);\\
g_0(t)=u(0,t),\quad h_0(t)=v(0,t);\\
g_1(t)=u_x(0,t),\quad h_1(t)=v_x(0,t).
\end{array}\label{slisp}\eeqa
Like Eq.(2.24), by using the initial and boundary data to define the spectral functions $s(\lambda)$ and $S(\lambda)$, we can further define the jump matrix $J_{m,n}(x,t,\lambda)$. Assume that the zero points of the  $(s^TS^A)_{11}(\lambda),s_{11}(\lambda),m_{11}(s)(\lambda)$ and $(S^Ts^A)_{11}(\lambda)$ are $\lambda_j(j=1,2\cdots N)$ just like in assumption 2.5. We also have the following results
\eqa\begin{array}{l}
u(x,t)=2i\lim_{\lambda\rightarrow\infty}(\lambda M(x,t,\lambda))_{12},\\
v(x,t)=2i\lim_{\lambda\rightarrow\infty}(\lambda M(x,t,\lambda))_{13}.
\end{array}\label{slisp}\eeqa
where $M(x,t,\lambda)$ satisfies the following $3\times 3$ matrix Riemann-Hilbert problem:

(1)$M$ is a sectionally meromorphic on the Riemann $\lambda$-sphere with jumps across the contours on $\bar D_n\cap\bar D_m(n,m=1,2,3,4)$ (see figure 2).

(2)$M$ satisfies the jump condition with jumps across the contours on $\bar D_n\cap\bar D_m(n,m=1,2,3,4)$
\beq M_n(\lambda)=M_mJ_{m,n},\quad \lambda\in \bar D_n\cap \bar D_m,n,m=1,2,3,4;n\neq m.\label{slisp}\eeq

(3)$M(x,t,\lambda)=\mathbb{I}+\mathcal{O}(\frac{1}{\lambda}),\quad \lambda\rightarrow\infty.$

(4)The residue condition of $M$ is showed in Proposition 2.6.
\end{thm}

Proof: We can use similar method with \cite{Xu2013} to prove this Theorem, It only remains to prove Eq.(3.2) and this equation follows from the large $\lambda$ asymptotic of the eigenfunctions.

\end{document}